# Coronal Flux Ropes and their Interplanetary Counterparts

N. Gopalswamy[a*], S. Akiyama[a,b], S. Yashiro[a,b], and H. Xie[a,b]

[a]NASA Goddard Space Flight Center Greenbelt, MD 20771, USA

[b]The Catholic University of America, Washington DC 20064, USA

*Corresponding author. e-mail: nat.gopalswamy@nasa.gov

**Abstract:**

We report on a study comparing coronal flux ropes inferred from eruption data with their interplanetary counterparts constructed from in situ data. The eruption data include the source-region magnetic field, post-eruption arcades, and coronal mass ejections (CMEs). Flux ropes were fit to the interplanetary CMEs (ICMEs) considered for the 2011 and 2012 Coordinated Data Analysis Workshops (CDAWs). We computed the total reconnected flux involved in each of the associated solar eruptions and found it to be closely related to flare properties, CME kinematics, and ICME properties. By fitting flux ropes to the white-light coronagraph data, we obtained the geometric properties of the flux ropes and added magnetic properties derived from the reconnected flux. We found that the CME magnetic field in the corona is significantly higher than the ambient magnetic field at a given heliocentric distance. The radial dependence of the flux-rope magnetic field strength is faster than that of the ambient magnetic field. The magnetic field strength of the coronal flux rope is also correlated with that in interplanetary flux ropes constructed from in situ data, and with the observed peak magnetic field strength in ICMEs. The physical reason for the observed correlation between the peak field strength in MCs is the higher magnetic field content in faster coronal flux ropes and ultimately the higher reconnected flux in the eruption region. The magnetic flux ropes constructed from the eruption data and coronagraph observations provide a realistic input that can be used by various models to predict the magnetic properties of ICMEs at Earth and other destination in the heliosphere.

Key words: coronal mass ejection, post eruption arcade, reconnected flux, flux rope

## 1. Introduction

Coronal mass ejections (CMEs) are one of the most important players in solar terrestrial relationship owing to their ability to cause intense geomagnetic storms and large solar energetic particle events (see e.g., Zhang et al. 2007; Gopalswamy 2009; 2010a). CMEs that continue into the heliosphere to become interplanetary CMEs (ICMEs) are on average fast and wide (Gopalswamy et al. 2010a). In other words, CMEs surviving far into the IP medium are generally more energetic. The energy of the CMEs can be traced to the free energy available in source magnetic regions on the Sun. Weak correlation between CME speed and the free energy available in the source regions has been reported (Gopalswamy et al. 2010a; Gopalswamy 2011) based on the assumption that the magnetic potential energy is a good proxy to the free energy (Mackay et al. 1997; Forbes 2000; Metcalf et al. 1995). A close connection between CMEs and flares is also expected based on the standard eruption model known as the Carmichael – Sturrock – Hirayama – Kopp and Pnueman (CSHKP) model (Carmichael 1964; Sturrock 1966, Hirayama 1974, Kopp and Pneuman 1976). The model involves the formation and ejection of a plasmoid with simultaneous formation of a post eruption arcade (PEA) due to RC. In three dimensions, the plasmoid is a flux rope (see e.g., Shibata et al. 1995; Longcope et al. 2007).

Soft X-ray emission from PEAs represents the flare, while the ejected flux rope represents the CME. Thus a close connection between CMEs and flares is expected except in confined flares, which do not involve any mass motion (Gopalswamy et al. 2009a). Correlating flare energy and CME kinetic energy with the active region potential energy neglects some key details of an eruption. For example, eruptions generally do not cover the entire active region area. It is well known that when there are multiple neutral lines present in an active region, eruptions can occur at different neutral lines at different times showing very different PEA morphologies. A well-known example is the two extreme events on 2003 October 28 and 29 that occurred on two different neutral lines in active region 10486. Accordingly, the PEA of the October 28 event formed over a horizontal neutral line, while the October 29 PEA was predominantly in the north-south direction because the near-vertical neutral line involved (Gopalswamy et al. 2005a; Gopalswamy 2008). Therefore, it important to consider only that section of the active region underlying the PEA. Secondly, the PEA "matures" in the decay phase of the flare, and hence incorporates the time evolution of the eruption. Thus an estimation of the total reconnected flux in the source region should be a better eruption characteristic that should be closely related to the flare fluence and the CME speed (or kinetic energy).

Longcope et al. (2007) showed that the PEA and flux rope are natural products of the reconnection process, so the poloidal flux of the flux rope should be the same as the reconnected (RC) flux in the source region. It is possible to test this flux relationship using solar and interplanetary data since the ejected flux ropes at the Sun are detected in situ in the solar wind as magnetic clouds (MCs) (Burlaga et al. 1981; Goldstein et al. 1983; Klein and Burlaga, 1982). Qiu et al. (2007) confirmed that the RC flux at the Sun is about the same as the poloidal flux of the corresponding MC at 1 au (see also Moestl et al. 2009; Hu et al. 2014; Gopalswamy et al. 2017). Qiu et al. (2007) determined the RC flux as the photospheric magnetic flux underlying the area swept up by the flare ribbons on one side of the polarity inversion line. Gopalswamy et al. (2017) introduced a new method of estimating the reconnected flux based on just PEAs, somewhat similar to the "flare flux" computed by Moore et al. (2007). According to this arcade method, a single image in EUV or X-ray in the decay phase of the flare can be used to estimate the RC flux: half of the magnetic flux underlying the PEA is the RC flux. They confirmed this by comparing the RC flux from both flare-ribbon and the arcade methods. They also confirmed the relation between the RC flux and the poloidal flux of the associated MCs for a larger sample of events than in Qiu et al. (2007).

One of the earliest ideas on CME flux ropes near the Sun was proposed by Mouchovias and Poland (1978). They successfully applied the flux rope model to a Skylab CME and found reasonable geometrical and magnetic properties of the flux rope. Chen et al. (1997) identified flux rope morphology in white-light CMEs observed by the Solar and Heliospheric Observatory (SOHO, Domingo et al. 1995) mission. Gibson and Low (1998) modeled CMEs with complex three-part structures with a 3D flux rope. Krall and St Cyr (2006) constructed a three-dimensional flux-rope model and found it to have good match with white-light observations. Krall (2007) further showed that CMEs can be modeled as hollow flux ropes. Thernisien (2011) introduced the graduated cylindrical shell (GCS) model to describe white-light CMEs observed in multiple views using SOHO and the Solar Terrestrial Relations Observatory (STEREO, Kaiser et al. 2008). Xie et al. (2013) applied the elliptical flux rope (EFR) model of Krall and St Cyr (2006) to show that a flux rope can be fit to white light CME irrespective of their appearance at 1 au as MCs or non-cloud ejecta (EJ).

The idea that all ICMEs have flux rope structure has been confirmed by many studies that show that both MCs and EJs contain flux ropes (Marubashi et al. 1997; Owens et al. 2005; Gopalswamy et al. 2013a; Xie et al. 2013; Makela et al. 2013; Yashiro et al. 2013; Kim et al. 2013; Marubashi et al. 2015). Numerical models use flux rope as a fundamental magnetic structure to model CMEs from the Sun to 1 AU (see e.g. Toth et al. 2007; Jin et al. 2017a). However, a flux rope may not be observed at 1 au when the observing spacecraft does not pass through the axis of the flux rope (Gopalswamy 2006). Such a situation is expected for CMEs originating at large central meridian distances. However, eruptions originating within a longitudinal distance of 15° also ended up as EJs at 1 au (Gopalswamy et al. 2013a). Examination of such events has revealed that the associated CMEs deflected away from the Sun-Earth line (Xie et al. 2013; Kim et al. 2013) because of the presence of coronal holes near the source regions (Mäkelä et al. 2013). Furthermore, the charge state properties of MC and EJ events were similar, suggesting that both were formed due to reconnection involving flares and flux ropes (Gopalswamy et al. 2013b). Both MC and EJ events were associated with similar PEAs suggesting that the eruption was similar producing PEAs and flux ropes (Yashiro et al. 2013). Finally, Marubashi et al. (2015) were able to fit flux ropes to even EJ events taking into account of the different impact parameters and adjusting the EJ boundaries in the solar wind data.

In this paper, we examine the RC flux in the solar sources of EJ and MC events and compare it with the flare, CME, and ICME properties. This study further confirms that the flux rope is a fundamental structure in all large solar eruptions and that the RC flux helps define the complete properties (geometrical and magnetic) of CME flux ropes near the Sun.

## 2. Data Selection

We make use of the list of 54 eruptions in solar cycle 23 considered for the Flux Rope CDAW workshops (Gopalswamy et al. 2010b; Gopalswamy et al. 2013b). Only eruptions occurring from within ±15° in longitude of the disk center were considered to make sure the CME reaches Earth to be detected by one or more of the Advanced Composition Explorer (ACE, Stone et al. 1998), Wind (Acuña et al. 1995), and SOHO spacecraft. The associated CMEs were observed by the Large Angle and Spectrometric Coronagraph (LASCO, Brueckner et al. 1995). Some ICMEs were consistent with the classical definition of MCs (enhanced field strength, smooth rotation of one of the components and low plasma temperature, see Klein and Burlaga 1982); those that did not agree with this definition were classified as EJs. For MCs, it was possible to fit a force-free cylindrical flux rope to the in situ solar wind plasma and magnetic field data (e.g., Lepping et al. 1990). The selected MCs can also be found on line with all the fitted parameters listed (http://wind.nasa.gov/mfi/mag_cloud_S1.html). Only 23 of the 54 ICMEs were well-defined MCs and the remaining 31 ICMEs were classified as EJs. However, Marubashi et al. (2015) have shown that it is possible to fit flux ropes even to EJ events. Therefore, we shall retain the EJ and MC labels as in the original list, with the understanding that both MC and EJ events have flux rope structure.

Our aim is to combine the source magnetic properties and the flux-rope geometric properties to fully describe the "magnetized" flux ropes in terms of the axial field strength and the poloidal and toroidal fluxes. In order to do this, we compute the RC flux from source regions using the PEA technique recently developed by Gopalswamy et al. (2017). According to this method, the RC flux is computed as half the unsigned photospheric magnetic flux underlying PEAs in the source region. The RC flux has been taken to be the poloidal flux of the flux rope created during the eruption process as has been demonstrated theoretically (Longcope et al. 2007) and observationally (Qiu et al. 2007; Hu et al. 2014; Gopalswamy et al. 2017). Equating the RC flux to the poloidal flux of the newly formed flux rope at the Sun, we determine the axial field in the coronal flux rope. In addition, we compare the near-Sun flux rope properties with near-Earth properties obtained from Marubashi et al. (2015) flux rope fits to in situ data.

Table 1 lists the properties of the 54 ICMEs from the CDAW list. The event identification number in column 1 corresponds to the original numbers used in the CDAW list (Gopalswamy et al. 2013a; Marubashi et al. 2015). The ICME type given in column 2 is based on the classical definition of magnetic clouds identified and modeled by the Wind Magnetic Field Investigation (MFI, Lepping et al., 1995) team. Other ICMEs are designated as non-cloud ejecta (EJ). The observed date, start time, and the peak magnetic field strength of the ICMEs are given in columns 3, 4, and 5, respectively. Properties of the associated CMEs are given columns 6-11 in the following order: CME date, time, sky-plane width, sky-plane speed ($V_{sky}$), space speed ($V_{sp}$), and mass. The space speeds were obtained from the halo CME list available at the CDAW Data Center (https://cdaw.gsfc.nasa.gov/CME_list/halo/halo.html). This list includes only full halo CMEs. For other CMEs, an empirical relation obtained by Gopalswamy et al. (2015a) were used: $V_{sp} = 1.10\ V_{sky} + 156$ in km/s. The CME mass estimates are as listed in the SOHO/LASCO CME catalog (https://cdaw.gsfc.nasa.gov/CME_list/index.html, Yashiro et al. 2004; Gopalswamy et al. 2009b). For 8 full halo CMEs, the mass estimate is not available. For these, we used an average value of $10^{16}$ g based on the fact that halo CMEs are generally fast, wide and massive (Gopalswamy et al. 2005b). Properties of the associated flares are listed in columns 12-15: heliographic coordinates of the flare location, GOES soft X-ray flare size, soft X-ray fluence, and the RC flux ($\Phi_r$). The fluence values were taken from the Solar Geophysical Data for most of the events; for some events (especially the ones associated with eruptive prominences) the fluence was not listed, so we computed it from the GOES soft X-ray flux. $\Phi_r$ was determined from the PEA method, which is half the sum of photospheric magnetic flux in each pixel in the area underlying the PEA (Gopalswamy et al. 2017). Xie et al. (2013) applied the EFR model to the white light images obtained by SOHO/LASCO using its C2 and C3 telescopes. This model assumes that the CME flux rope has an elliptical axis with varying radial circular cross-section. The aspect ratio is the main geometrical parameter related $\Lambda$, the ratio of the heliocentric distance to the radius ($R_0$) of the flux rope at its apex. In terms of the heliocentric distance ($R_{tip}$) to the tip of the flux rope, $\Lambda = ½ (R_{tip} - R_0)/R_0$. The $\Lambda$ parameter obtained from the EFR fits is given in column 16 while the $R_{tip}$ values used in computing $\Lambda$ are given in column 17. The flux rope fit also gives the direction of the flux rope (column 18), which can be different from the flare location because of non-radial motion of the flux rope (Gopalswamy et al. 2009c; Xie et al. 2013; Makela et al. 2013; Gopalswamy et al. 2014a). By tracking the leading edge of the flux rope, we also get the deprojected speed of the flux rope ($V_{spf}$) given in column 18 giving another estimate of the space speed listed in column 10. Under the assumption of self-similar expansion and the known result that the flare RC flux is approximately the same as the poloidal flux of the flux rope formed due to reconnection, we obtain the axial magnetic field strength of the flux rope in the corona as $B_0 = \Phi_r x_{01}/LR_0$, where $x_{01}$ is the first zero of the Bessel function $J_0$ and $L$ is the length of the flux rope taken as $2R_{tip}$. We can calculate $B_0$ at any distance, but we have listed the value at 10 Rs in column 20 in units of mG. Marubashi et al. (2015) fitted flux ropes to the 1-au data of to both MC and EJ events giving flux rope parameters such as the axial field strength $B_0$ and radius $R_0$ at 1 au. From these, we determined the poloidal flux of the 1-au flux ropes as $\Phi_p$ (1 au) = $LB_0 R_0/x_{01}$. We assumed the flux rope length $L$ to be 2 au at Earth. In four ICMEs Marubashi et al. (2015) had to split the ICMEs into more than one flux rope to do the fits. We excluded these ICMEs in this study. The resulting $\Phi_p$ (1 au) and $B_0$ (1 au) are listed in columns 21 and 22, respectively. We use the data in Table 1 to relate the eruption data to the coronal and interplanetary flux rope properties.

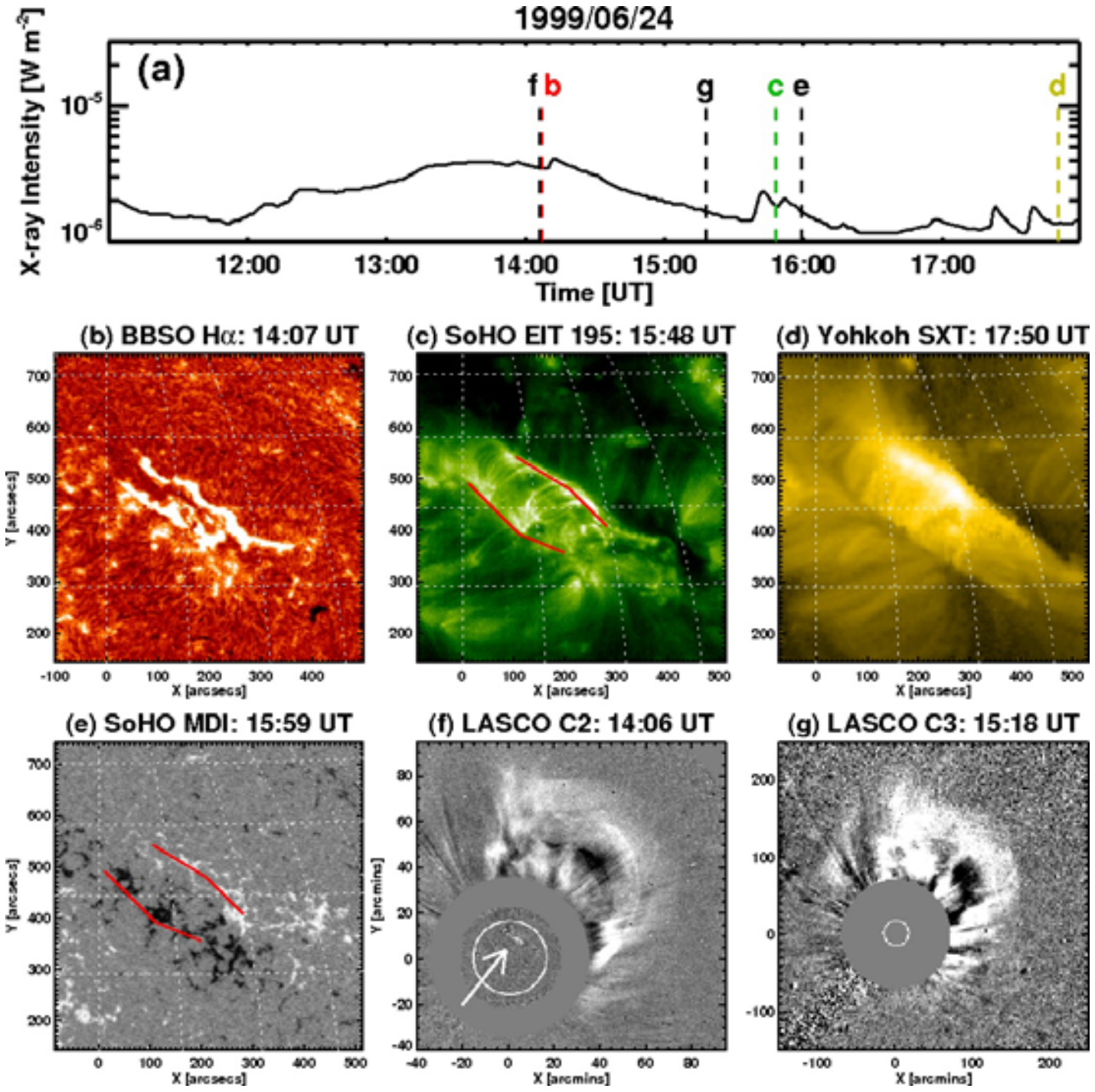

*Figure 1. The solar source, flare, and CME in the 1999 June 24 eruption. (a) GOES soft x-ray light curve in the 1-8 Å band showing the C4.1 gradual flare associated with the eruption. (b) H-alpha picture from the Big Bear Solar Observatory (BBSO) showing the two ribbon flare. (c) The post eruption arcade (PEA) observed in EUV by SOHO/EIT in the 195 Å waveband with the footpoints of the arcade marked by the red lines, (d) The PEA observed Yohkoh's Soft X-ray Telescope (SXT), (e) the PEA footpoints superposed on a SOHO/MDI magnetogram taken at 15:59 UT, (f) a SOHO/LASCO difference image at 14:06 UT with a EIT 195 Å difference image superposed showing the CME and the solar source (pointed by arrow), and (g) a*

*SOHO/LASCO/C3 difference showing that the CME became a full halo. The times of various images are marked on the GOES plot in (a) (not in chronological order).*

**3. RC Flux and Flare Properties**

Figure 1 shows the solar source of one of the events in Table 1 (#10, 1999 June 24). The eruption was characterized by a two-ribbon flare observed in H-alpha and a PEA centered at N29W13 observed by Yohkoh's Soft X-ray Telescope (SXT, Tsuneta et al. 1991) and SOHO's Extreme-ultraviolet Imaging Telescope (EIT, Delaboudiniere et al. 1995). The PEA was quite extended in the northeast-southwest direction and was formed due to the eruption of a filament (also reported as a disappearing solar filament event from N33E09 during 10:51 to 14:18 UT in the Solar Geophysical Data). In soft X-rays, the flare was of C4.1 class observed by GOES and had a fluence of $3.3\times10^{-2}$ J m$^{-2}$. The eruptive filament became the core of a large CME observed by SOHO/LASCO. The CME was relatively fast with an average speed of ~975 km/s in the SOHO/LASCO field of view (FOV) and showed a positive acceleration (32.5 m s$^{-2}$) in the corona, typical of CMEs associated with quiescent filament eruption (Gopalswamy et al. 2015b). By the time the CME reached the edge of the LASCO FOV, it had a speed of 1200 km/s. The CME was a partial halo in the C2 FOV and became a full halo in the C3 FOV. When projection corrected using the cone model (Xie et al. 2004), the CME had an average speed of 1143 km/s. From the leading edge measurements of the flux rope obtained from the EFR model, the space speed was ~1543 km/s. The aspect ratio $\Lambda$ was 0.70 at an $R_{tip}$ of 7.5 Rs. The flux rope direction (N25W15) was slightly different from the direction indicated by the flare location.

The PEA was superposed on a magnetogram obtained by the Michelson Doppler Imager (MDI, Scherrer et al. 1995) at 15:59 UT to get the RC flux of $5.2\times10^{21}$ Mx. The ICME was an EJ event arriving at Sun-Earth L1 on 1999 June 27 at 21:30 UT. The fitted value of $B_0$ from Marubashi et al. (2015) was ~8.7 nT and $\Phi_p$ (1 au) ~ $0.77\times10^{21}$ Mx. The observed maximum value of the ICME magnetic field was ~12.1 nT. All these parameters compiled for this event can be found in

Table 1. Comparing the PEA between EIT and SXT images, we see that the SXT arcade has an extension to the south, which is not clear in the EIT image. When we include the additional area seen in SXT image, we get an RC flux of $7.6\times10^{21}$ Mx, which is larger than the value obtained from EIT by 46%. We decided to stay with the EIT data for consistency because we do not have SXT observations for all the events.

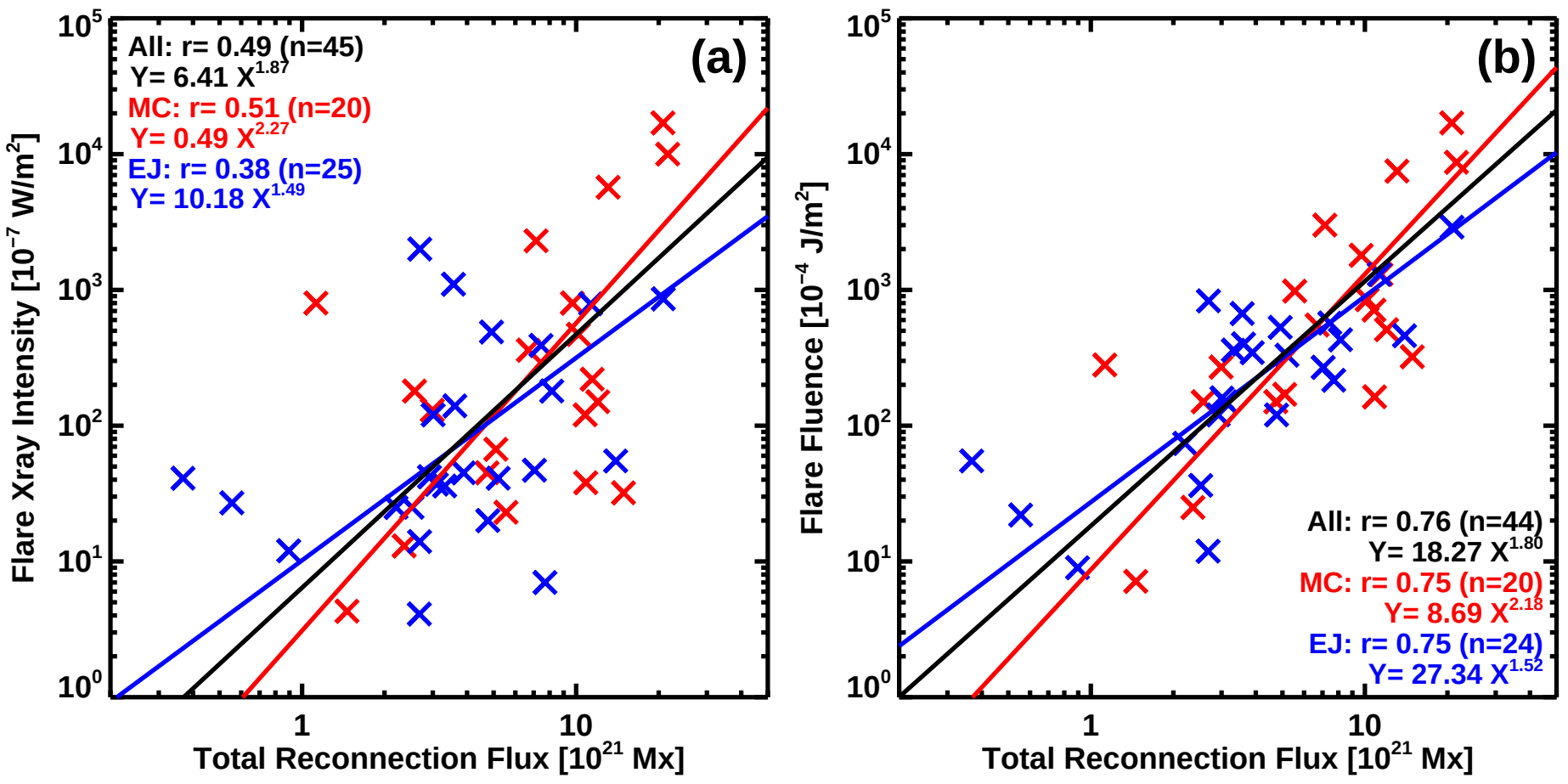

*Figure 2. Scatter plot between the RC flux and (a) the soft X-ray flare size and (b) the soft X-ray flare fluence. The red and blue data points (with the corresponding regression lines) denote flares associated with MCs and EJs, respectively at 1 au. The black line is the regression of the combined data set ("All"). This color-coding is used throughout this paper to distinguish the MC, EJ, and combined data set. The correlation coefficients (r) and the number of data points (n) are shown on the plots. In the RC flux – flare size correlation, the Pearson's critical correlation coefficients ($r_c$) at 95% confidence level are: 0.251 (All), 0.389 (MC), and 0.337 (EJ); in the RC flux – flare fluence correlation, the critical coefficients at 95% confidence level are: 0.254 (all), 0.389 (MC), and 0. 344 (EJ). In both cases, the probability that the correlations for the combined set is by chance is less than $5\times10^{-4}$. The $r_c$ values quoted throughout this paper correspond to 95% confidence level.*

### 3.1 The RC flux, Flare Size, and Flare Fluence

The RC flux from a source region depends on the total amount of magnetic flux involved in an eruption. Observationally, the RC flux is given by the area swept up by the flare ribbons (or the PEA) on one side of the polarity inversion line in the source region and the photospheric magnetic flux under this area. On the other hand the peak soft X-ray flux and the fluence are due to the hot flare plasma generated as a result of electrons accelerated during the reconnection and precipitating in the solar atmosphere. Therefore, we do expect that the RC flux is related to the flare intensity and fluence. Figure 2 shows this relationship for all the events in Table 1 that have these measurements. We see that both the flare size and fluence have high correlation with the RC flux. The fluence and flare size vary by more than three orders of magnitude, while the RC flux varies over two orders of magnitude. There is very good overlap between MC and EJ events, suggesting that there is no notable difference in fluences and flare sizes of the two ICME populations. The fluence – RC flux correlation is much stronger because both quantities refer to the total duration of the eruption, while the flare size corresponds to one instance without considering the rise and decay phases.

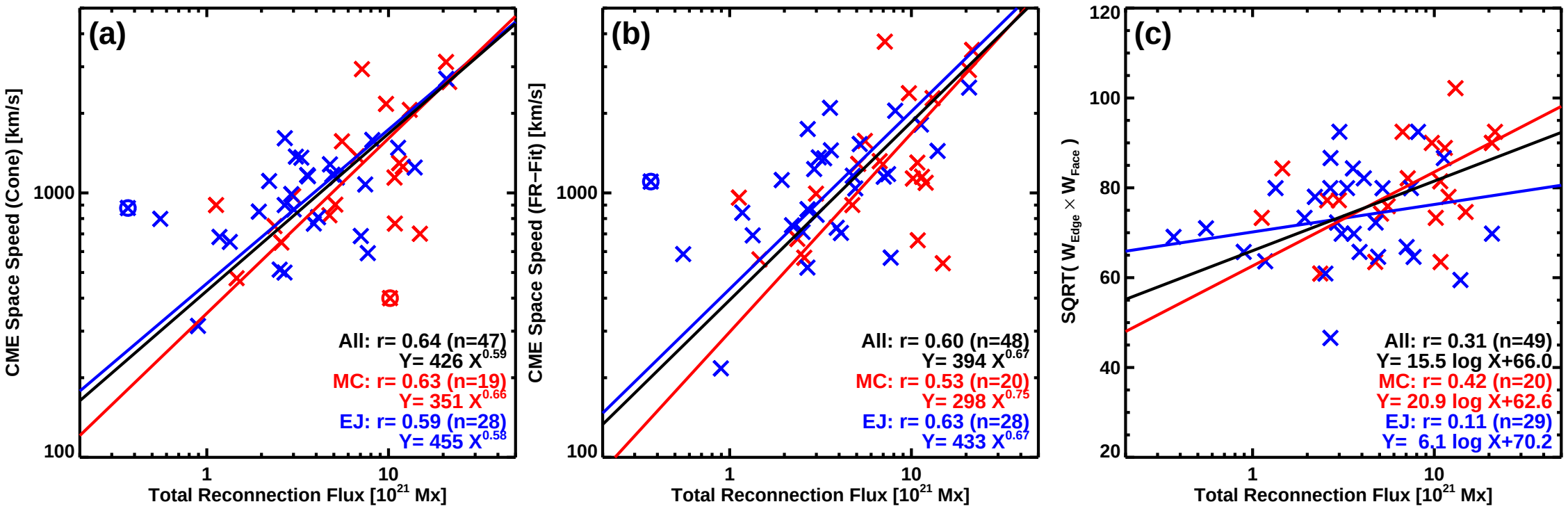

*Figure 3. Scatter plot between CME speed and RC flux for MC (red) and EJ (blue) events: (a) when space speed was determined by correcting the sky-plane speed for projection effects using the cone model and (b) when space speed was obtained by tracking the leading edge of the flux rope. In (a) and (b) event #25 (2000 October 2) is excluded from the correlations because the arcade is ill-defined. Event #43 (2002 July 29) is excluded from (a) because of the large discrepancy between the cone model speed (400 km/s) and the flux-rope speed (1134 km/s). It is*

*possible that the LASCO speed measurement is incorrect. The two excluded data points are shown circled. (c) Scatter plot between the RC flux and the flux rope width obtained as a geometrical mean of the edge-on and face on width (see Xie et al. 2013). All correlations are significant because the Pearson critical values of the correlation coefficients are much smaller.*

### 3.2. The RC Flux and CME Properties

In an earlier work connecting CME widths and the flare flux, Moore et al. (2007) showed that the CME width is decided by the average magnetic field under the flare arcade. Since the RC flux is proportional to the average field strength under the arcade, we do expect a correlation between the RC flux and CME width. However, there is a problem in measuring the CME widths for the events in Table 1: the CMEs are all disk-centered, so most of them are halo CMEs and hence their widths are known from single-view observations. Since the white-light CMEs were fit to a flux rope, we do have the edge-on ($W_{edge}$) and face-on ($W_{face}$) widths of the CME flux ropes. We use the quantity $(W_{edge} \times W_{face})^{1/2}$ as a measure of the CME width. We do have good CME speed measurements in the sky plane. As noted in section 2, we do have space speeds from cone model deprojection and flux rope tracking.

Figure 3 shows that the CME speed is well correlated with the RC flux for both EJ and MC events suggesting that one cannot easily distinguish the two populations based on the CME speeds. We have shown the correlations using space speeds obtained by the cone-model deprojection (Fig. 3a) and flux-rope tracking methods (Fig. 3b). The correlations in Figs. 3a and 3b are quite similar because the two speeds agree quite well. In Fig. 3c, we have shown the correlation between the RC flux and the flux-rope width defined as $(W_{edge} \times W_{face})^{1/2}$. We see a significant correlation for all events ($r$ =0.31, $r_c$ = 0.238) and for MC events ($r$ = 0.42, $r_c$ = 0.378). The correlation is also positive in the case of EJ events, but it is not statistically significant.

The correlation of RC flux with CME speed and width implies that it should also be correlated with the CME kinetic energy. It is well known that the CME mass (M) is related to CME width (W) according to: log M = 12.6 log W (Gopalswamy et al. 2005b). Furthermore, the CME width is also correlated with CME speed (Gopalswamy et al. 2014b). Accordingly, we expect a better correlation between RC flux and CME kinetic energy. Mass estimates are available only for 40 events, which we combined with the space speed from the two methods (cone model and flux rope fit) to get kinetic energies. The SOHO/LASCO catalog lists mass values for CME widths in the range 20-120 degrees. For nine events, we assumed an average mass of $10^{16}$ g based on the fact that CMEs associated with ICMEs are generally faster and wider on average and a majority of them are halo CMEs. Furthermore, a compilation of masses of limb CMEs indicates that the bin with ~$10^{16}$ g corresponds to above average values (Gopalswamy, 2010a).

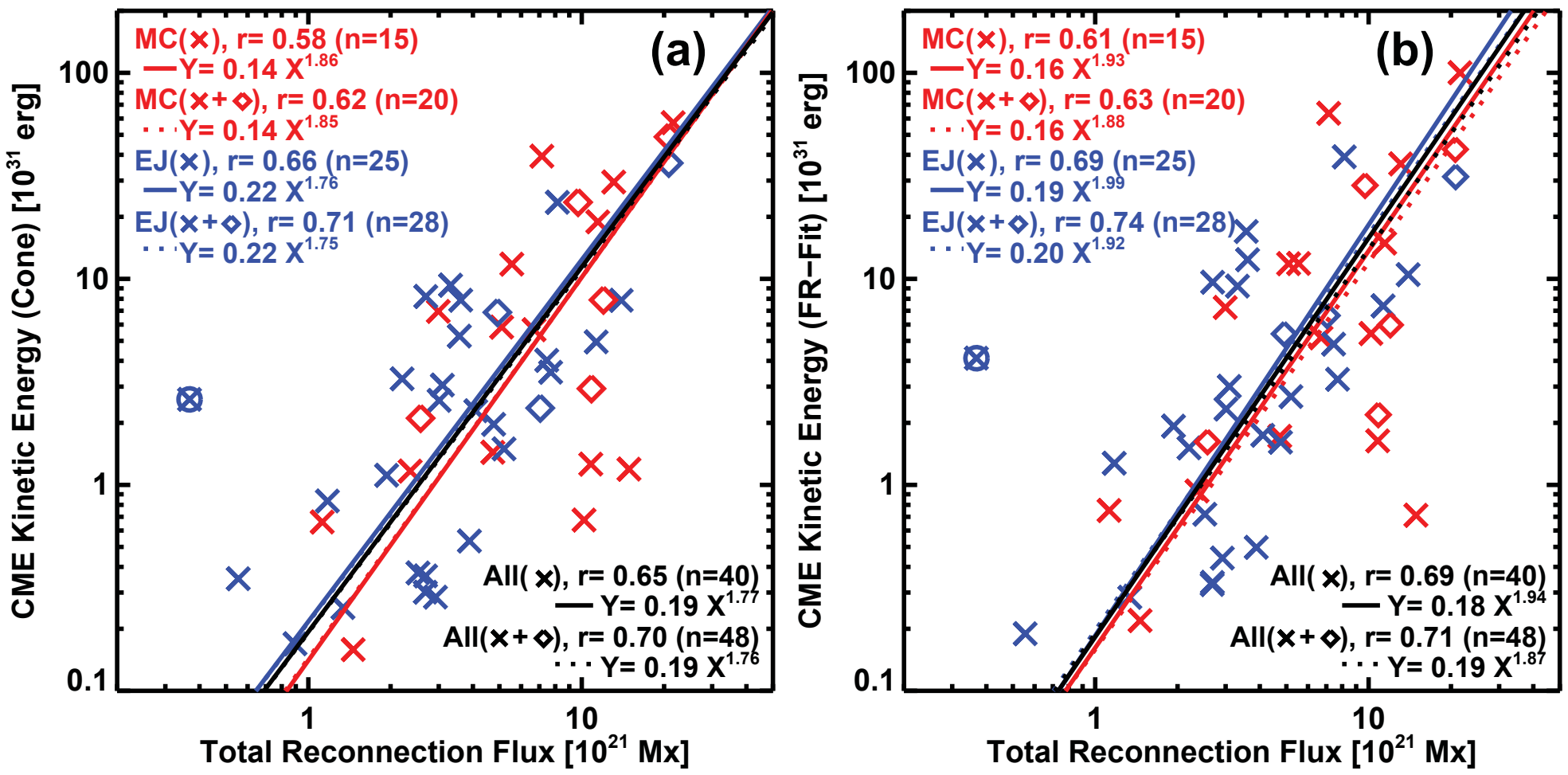

*Figure 4. Scatter plot between CME kinetic energy and RC flux. The kinetic energy was calculated from the CME mass listed in the SOHO/LASCO catalog and the space speed derived from the cone model deprojection (a) or from flux rope fits to the white-light CME data (b). The kinetic energy values obtained by assuming CME mass ($10^{16}$ g for 5 MC events and 3 EJ events) are denoted by diamond symbols. Correlation coefficients and regression lines are shown with and without the few assumed masses. The regression lines are all very close to each other. All the correlation coefficients are much larger than the Pearson correlation coefficient and hence*

*are statistically significant. Event #25 is excluded (data point shown circled) for the reason stated in Fig. 3.*

The scatter plot between RC flux and CME kinetic energy in Fig. 4 shows that the two quantities have a better correlation than the RC flux –speed correlation. Comparing the kinetic energies obtained from the cone-model and flux-rope tracking speeds, we see that the correlation from the flux-rope tracking is slightly better. The correlations are significant for EJs ($r = 0.74$, $r_c = 0.317$ for 28 events), MCs ($r = 0.63$, $r_c = 0.378$ for 20 events) and the combined set ($r = 0.71$, $r_c = 0.243$ for 48 events) at 95% confidence level. These numbers include the 8 events for which the mass was assumed. Inclusion of the assumed-mass events did not change the correlation significantly, but does indicate consistency with the measured masses.

## 4. Flux Rope Magnetic Field in the Corona and IP medium

In this section we consider the magnetic field strength in the coronal and in the 1-au flux ropes using independent measurements. We assume force-free flux ropes in both cases. The 1-au flux rope parameters are listed in Marubashi et al. (2015). The magnetic field strength in coronal flux ropes is obtained from a combination of EFR model for geometrical properties and the RC flux computation using the arcade method. We also compare the properties of coronal flux ropes from the source regions of MC and EJ events.

### 4.1 CME Magnetic Field Strength in Coronal Flux Ropes

One of the immediate outcomes of the close relation between the RC flux and the poloidal flux of the erupted flux rope is that we can obtain the axial field strength ($B_0$) of the CME flux rope. Using the Lundquist solution for a force free flux rope, we get $B_0 = \Phi_r x_{01}/R_0 L = \Phi_r x_{01}/2R^2_0(2\Lambda+1)$. Here we have assumed that the poloidal flux of the flux rope in the corona equals the RC flux $\Phi_r$. Figures 5 (a,b) show histograms of the RC flux in the source regions of MCs and EJs; the distributions have different shapes, even though the range of values is similar.

The RC flux in the MC source regions is almost twice that in EJ source regions on average. Figures 5 (c,d) show the radius ($R_0$) of the coronal flux rope, which has similar distributions for MC- and EJ-associated CMEs. The distributions of the aspect ratio parameter $\Lambda$ is also similar for MC and EJ cases (not shown). The axial magnetic field strength $B_0$ of the white-light flux ropes derived from the RC flux and the fitted flux rope parameters is given in Figs. 5(e,f). We see that the coronal flux rope $B_0$ values are in MC and EJ events: the average $B_0$ of MC-associated CME flux ropes is about 65% higher than that in EJ-associated ones. Kolmogorov-Smirnov tests confirmed that distributions of both the RC flux and $B_0$ have statistically significant differences between MC and EJ events, while the difference in $R_0$ (and $\Lambda$) is not significant. Similar $R_0$ between MC and EJ events implies that the different $\Phi_r$ should lead to different $B_0$ because $\Phi_r \sim B_0R_0$ and $R_0 \sim$ constant between MC and EJ events for a given heliocentric distance.

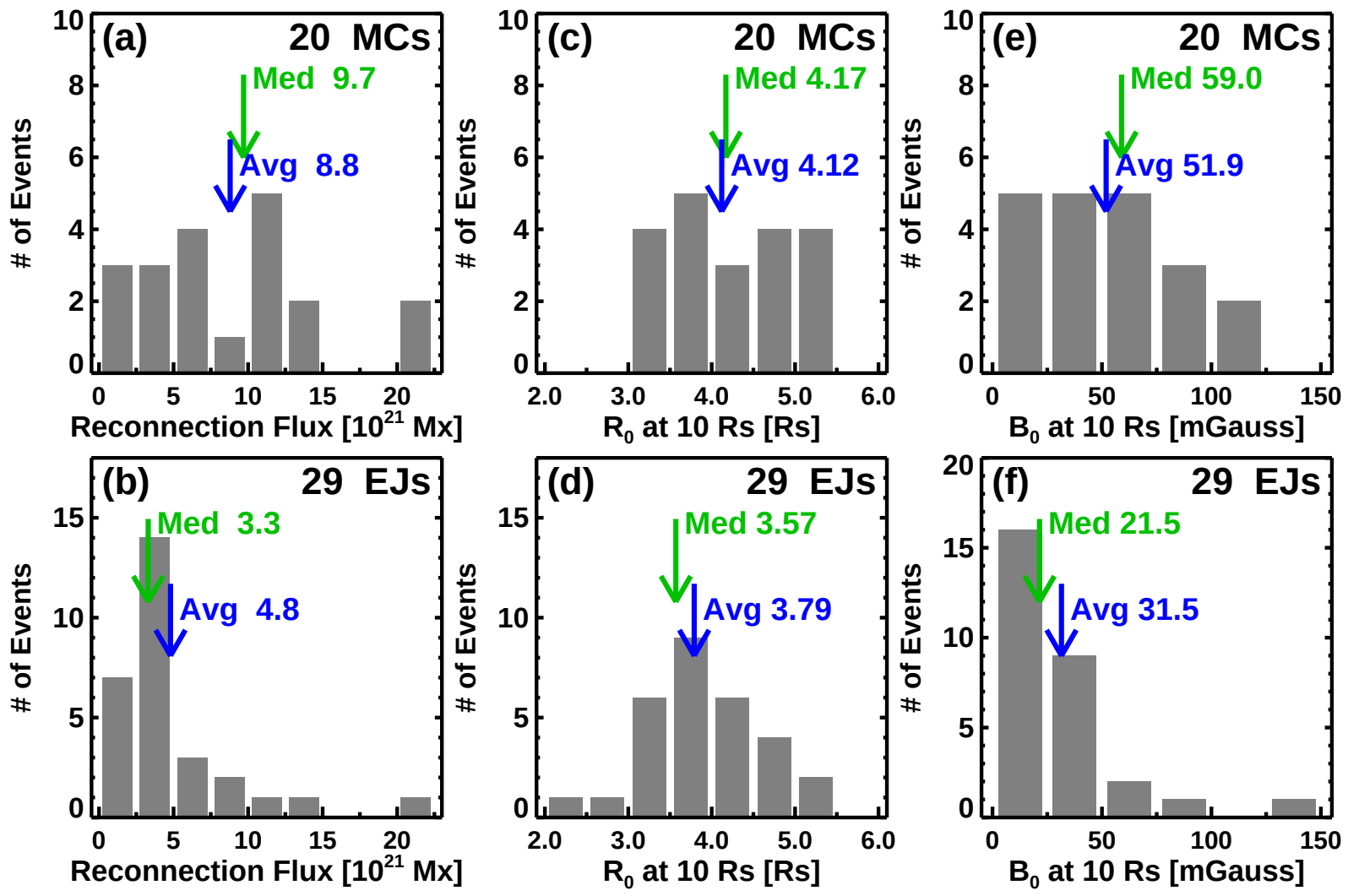

*Figure 5. Distributions of the RC flux $\Phi_r$ (a,b), the flux-rope radius $R_0$ (c,d), and the derived axial magnetic field strength $B_0$ of the flux rope at a heliocentric distance of 10 Rs (e,f). The top and bottom panels show the quantities associated with MCs and EJs, respectively. The mean (Avg) and median (Med) values are indicated on the plots. Kolmogorov-Smirnov Test comparing the MC and EJ parameters yielded the D-statistic as 0.4241 (RC flux), 0. 3017 ($R_0$) and 0.4741 ($B_0$). The corresponding probabilities (p) that the obtained D value is by chance are 0.019, 0.191,*

*and 0.006, indicating that $R_0$ has a similar distribution in MCs and EJs, while the RC flux and $B_0$ have different distributions.*

In an early work on CME flux ropes, Mouschovias and Poland (1978) showed that the thickness of the flux rope at the leading edge (and hence the flux rope radius) is proportional to the heliocentric distance of the flux rope ($R_0 \propto R_{tip}$). Since $L$=2 $R_{tip}$, we see that $B_0 \sim R^{-2}_{tip}$. This inverse square dependence on distance is faster than that for the background magnetic field in the corona obtained from various techniques. Gopalswamy and Yashiro (2011) determined the variation of the background magnetic field as a function of radial distance from shock standoff distance measured in white-light coronagraph images. They found a relationship $B(R) = 0.356\ R^{-1.28}$ for an adiabatic index of 5/3. For $R$ = 10 Rs, this relation yields a $B$ = 18.7 mG. Clearly, the coronal flux rope has a higher magnetic field strength (by a factor of 2.8 for MCs and 1.8 for EJs) than in the background corona. This is an important confirmation of the flux-rope nature of CMEs as a low-beta plasma throughout the inner heliosphere. The faster dependence of the flux-rope magnetic field compared to that in the background medium was also found in the specific event investigated by Mouschovias and Poland (1978). Here we are able to confirm that result statistically.

**4.2 Flux Rope Magnetic Field Strengths at the Sun and at 1 au**

Two-point measurements (one near the Sun by remote-sensing and the other near Earth by in situ measurements) of solar disturbances have been extremely useful in understanding the propagation of CMEs (Lindsay et al. 1999; Gopalswamy et al. 2000; 2001). Such studies mainly deal with the arrival time and speed of CMEs at 1 au based on CME kinematics at the Sun. A more realistic comparison requires predicting the magnetic properties of ICMEs, which are important for space weather applications. For example, a knowledge of the out-of-the ecliptic component of the magnetic field in ICMEs is critical in predicting the strength of geomagnetic storms (e.g., Wilson 1987; Gonzalez et al., 1987; Wu and Lepping, 2002; Zhang et al. 2007; Gopalswamy et al. 2008; Gopalswamy 2010b). The flux-rope nature of ICMEs, if known ahead of time, will tell us when and where the southward component of the magnetic field would occur; the product of the southward field component and the ICME speed is the critical in predicting the strength of the resulting geomagnetic storms. The close relation between the RC flux in the source region and the poloidal flux of 1-au MCs thus provides the required link between geomagnetic storms and solar source regions and supports data-constrained CME models (Jin et al. 2017b). We showed in Figs. 3 and 4 that the RC flux has good correlation with CME speed and kinetic energy. We now show that the RC flux has also a reasonable correlation with the 1-au poloidal flux for both MC and EJ events.

Since Marubashi et al. (2015) were able to fit flux ropes to both MC and EJ events, we use the fitted parameters (axial field strength $B_0$ and the flux rope radius $R_0$ at 1 au) of the flux ropes to obtain the poloidal flux at 1 au as $(L/x_{01})B_0R_0$. Here we assumed the length of the flux rope to be 2 au as in Gopalswamy et al. (2017). As we noted before, we consider only those events that were not split into multiple ICMEs in fitting flux ropes. There are 46 such events, whose 1-au properties are shown in Fig. 6. We see that $B_0$ of the 1-au flux ropes is higher in MCs by a factor of ~1.5 than in EJs (Fig. 6c), similar to what was found for coronal flux ropes. The poloidal flux derived from the 1-au fits also differ between MC and EJ events (Fig. 6a): the MC poloidal flux is ~73% higher on average. However, the radial size of the flux ropes are not too different. We confirmed that the differences in $B_0$ and poloidal flux between MC and EJ events are statistically significant, while the difference in $R_0$ is marginal. The EJ-type ICMEs seem to have flux ropes with smaller magnetic content (lower axial field strength and poloidal flux) compared to MCs. Thus we conclude that there are significant differences between the flux ropes underlying MC and EJ events, the root cause being the different RC fluxes. We also notice that the difference in $R_0$ at 1 au between EJ and MC is slightly more pronounced than in the corona. This is consistent with the higher magnetic content of MC flux ropes (and hence higher magnetic pressure) resulting in greater expansion and hence relatively a larger size at 1 au.

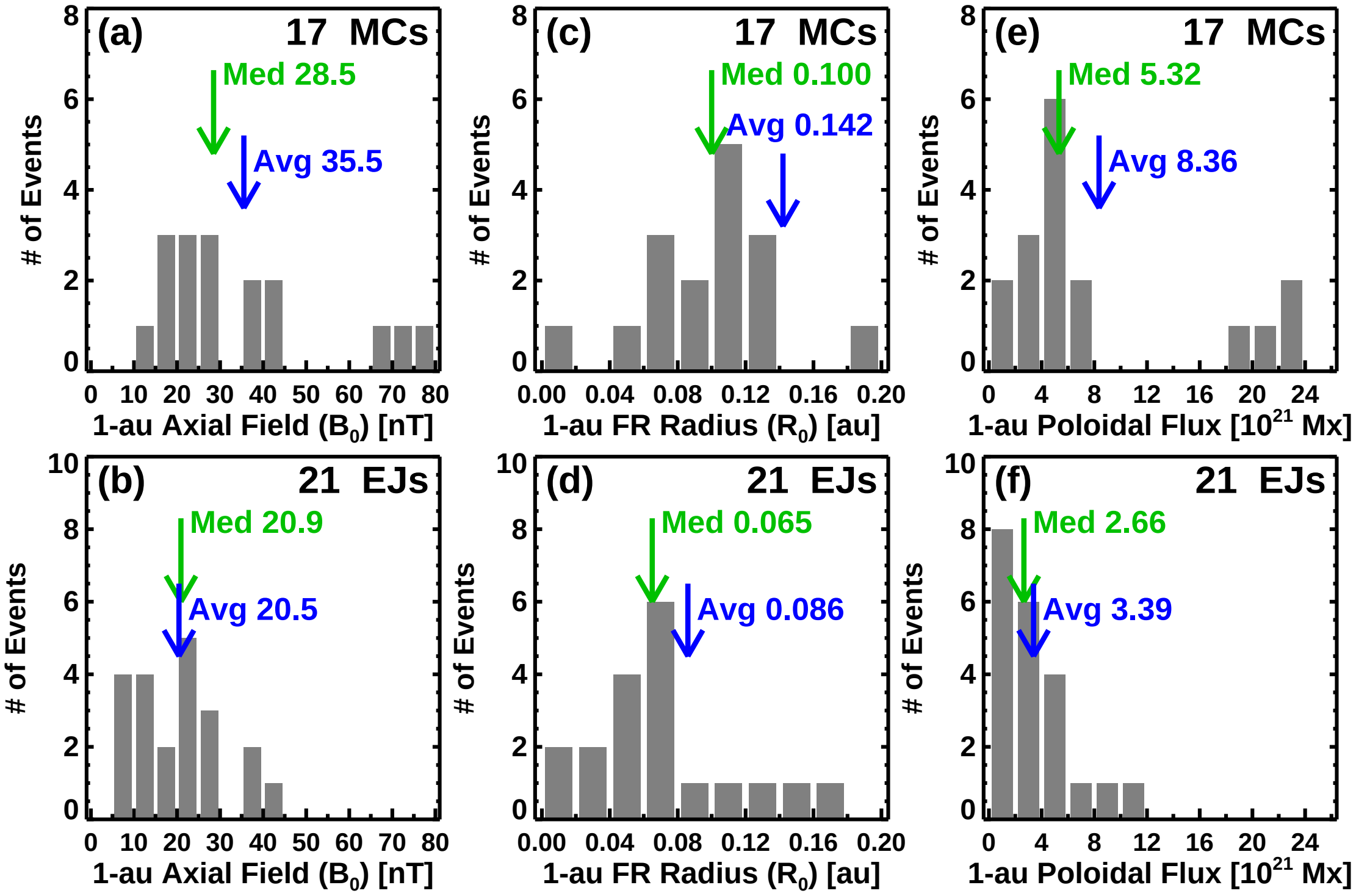

*Figure 6. Distributions of axial magnetic field strength (a,b), the flux-rope (FR) radius (c,d) and the derived flux-rope poloidal flux (e,f) at 1 au. The top and bottom panels show the quantities associated with MCs and EJs, respectively. The mean (Avg) and median (Med) values are indicated on the plots. Kolmogorov-Smirnov test comparing the MC and EJ distributions shows that the difference between MC and EJ events are significant for the axial field strength (a, b) and the poloidal flux (e,f). The difference is marginal in the case of the radius of the flux rope. The D-statistic for the three parameters (left to right) and the probability (p) that the D value is by chance are: 0.4174 (p=0.05), 0.3950 (p=0.08), and 0.4314 (p=0.04).*

Figure 7(a) shows the relation between the RC flux and the poloidal flux at 1 au for the events in Table 1; the correlation is significant for the combined data set ($r = 0.47$ with $r_c = 0.275$ at 95% confidence level). When MCs and EJs are considered separately, the correlation is still positive but not statistically significant because of the small sample sizes. The regression line for all events [$\Phi_p$(1 au) = 0.79($\Phi_r$)$^{1.04}$] is very close to the equal fluxes line. For $\Phi_r$ =10.0×10$^{21}$ Mx, the regression line gives $\Phi_p$ =8.7×10$^{21}$ Mx, which is 13% smaller. The regression line for MCs (based on Lepping et al. 1990 fits) obtained in Gopalswamy et al. (2017) was 1.20($\Phi_r$)$^{0.85}$, which yields a similar value: $\Phi_p$ = 8.5×10$^{21}$ Mx; the MC regression line (Fig. 7a) gives the same $\Phi_p$.

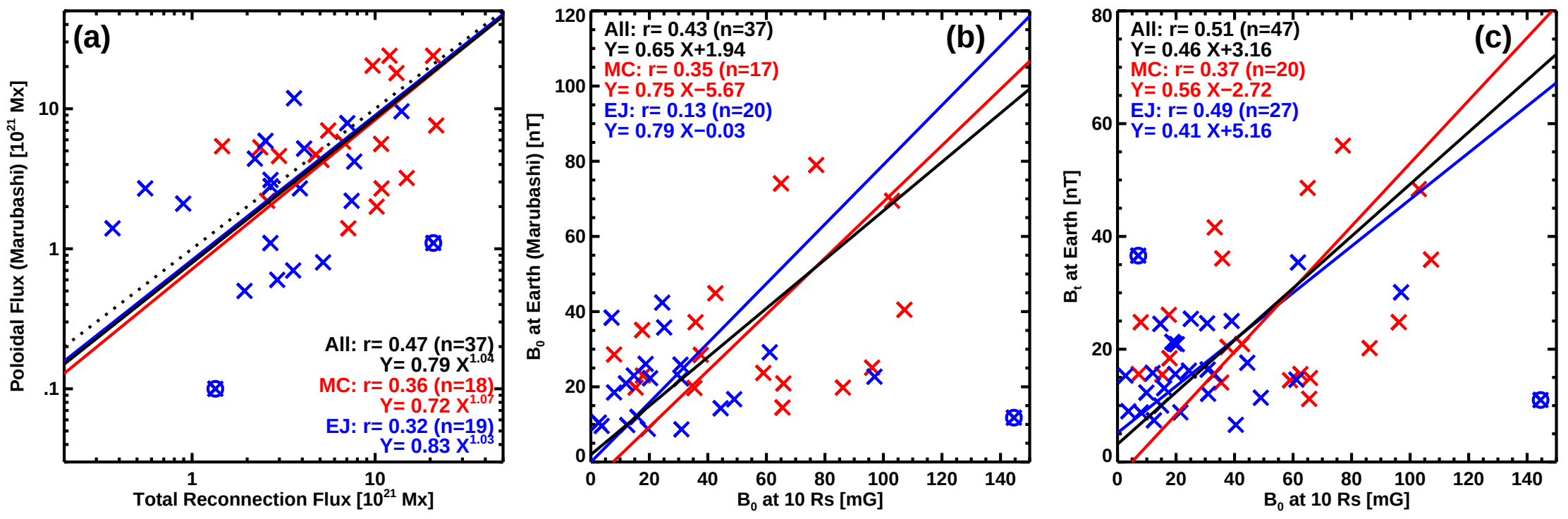

*Figure 7. (a) Scatter plot showing how the poloidal flux of the 1-au flux rope relates to the RC flux at the Sun. The combined set has a positive correlation (r = 0.47), which is statistically significant ($r_c$ = 0.275). The probability that the correlation is by chance is less than 0.005. Event #30 (2001 March 19 located at the lower-left part of the plot) excluded because Marubashi et al. (2015) subdivided the ICME interval into three events. Event #51 (located at the upper-right part of the plot) was also excluded because the ICME boundary used by Marubashi et al. (2015) was very different from that in the CDAW list. The outliers are shown circled. Including them reduced the correlation, but remained significant for the combined set. The dotted line represents equal fluxes on the X and Y axes. (b) Scatter plot between the axial field strength (in units of nT) obtained from flux rope fit to the in situ data (Marubashi et al. 2015) and the $B_0$ obtained at 10 Rs (in units of mG) by fitting a flux rope to the white-light CME data. The correlation is again significant for the combined data set (r = 0.43 compared to $r_c$ = 0.275). (c) The correlation between the B0 at 10 Rs and the observed maximum value of the ICME magnetic field. The probability that the observed good correlation (r = 0.51 with $r_c$ = 0.243 for 47 events at 95% confidence level) is by chance is <5×10$^{-4}$. The $B_t$ value in event #14 (1999 October 21 EJ) is unusually high because of the compression by a CIR behind the ICME,*

*so it is excluded in the correlation. Event #51 was also excluded in (b) and (c) for the same reason given in (a).*

In Figure 7b, we show how the axial magnetic field $B_0$ at 10 Rs is related to that in the corresponding 1-au flux rope. The combined data set (MC + EJ) shows a reasonable correlation ($r = 0.43$ with $r_c = 0.275$). Taken separately, both MCs and EJs show positive correlation, but the correlations are not statistically significant due to insufficient sample sizes. It is certainly remarkable that quantities derived from completely independent observations made 1 au apart show reasonable correlation. Finally, Fig. 7c shows the correlation between the derived $B_0$ at 10 Rs and the observed maximum magnetic field strength ($B_t$) in the corresponding ICMEs. The correlation ($r = 0.51$ with $r_c = 0.243$) is certainly better than that in Fig. 7a,b. The high correlation is promising that the solar eruption data have a high prediction value not only for the arrival time, but also for 1-au magnetic properties, crucial for space weather applications.

## 5. Discussion

We characterized the dimensions and magnetic content of CME flux ropes near the Sun based on eruption data for a set of more than 50 events. We obtained the geometric properties of CME flux ropes (aspect ratio and the flux rope radius at the apex) by fitting a flux rope to the white-light coronagraphic observations from SOHO/LASCO. Since we had only single view observations during solar cycle 23, we used the EFR model of Krall and St Cyr (2006). We obtained the magnetic properties of the flux ropes (poloidal flux and axial field strength) using the RC flux in the eruption region; the RC flux itself was determined from the photospheric magnetic flux underlying the post eruption arcades (Gopalswamy et al. 2017). Other properties such as the helicity sign of the flux rope can easily be obtained from the hemispheric rule or using one of a number signatures in the source region, including the skew of the post eruption arcade (e.g., Bothmer and Schwenn 1998; Yurchyshyn et al. 2001; Yurchyshyn 2008; Sheeley et al. 2013). For force-free flux ropes, the toroidal flux can be readily obtained from the poloidal flux. Furthermore, one can also obtain the poloidal ($B_p$) and toroidal ($B_a$) components of the magnetic field strength from the Lundquist solutions (see e.g., Lepping et al. 1990): $B_a = B_0J_0(\alpha\rho)$, and $B_p = HB_0J_1(\alpha\rho)$, where $\alpha$ is the force-free parameter (constant), $\rho$ is the radial distance from the flux rope axis, and $H = \pm 1$ is the helicity sign. Thus, we have a complete description of the near-Sun flux rope that can be used in global MHD models to get asymptotic properties of flux ropes such as the magnetic field components and flux rope arrival times at various destinations in the heliosphere. The force-free flux rope assumption can also be tested using actual observations to make further improvements. Since we need to identify the post eruption arcade during the decay phase of flares, it will take a couple of hours to measure the RC flux once the eruption happens.

Throughout this work, we tacitly assumed that the CME flux rope is formed due to reconnection. The correlation between the RC flux and the poloidal flux of ICMEs shown in Fig. 7a justifies this assumption. The regression line is very close to the equal fluxes line in Fig. 7a, but there is a slight offset suggesting that the poloidal flux is slightly smaller than RC flux. If reconnection adds more flux to a pre-existing flux rope, one would expect the poloidal flux to exceed the RC flux. It appears that the pre-eruptive flux rope may possess only a small fraction of the post-eruptive flux rope. Another possible way in which the poloidal flux may decrease is erosion by interchange reconnection with the ambient magnetic field leading to an unbalanced flux rope configuration (e.g., Manchester et al. 2014). Currently, the extent of this process is not fully estimated. In a minority of MCs, an erosion of up to 40% has been reported (Ruffenach et al. 2015). Significant flux rope erosion will lead to the condition, $\Phi_r >> \Phi_p$ which is generally not the case. Therefore, we assume that the erosion is not significant, taken all the MCs together.

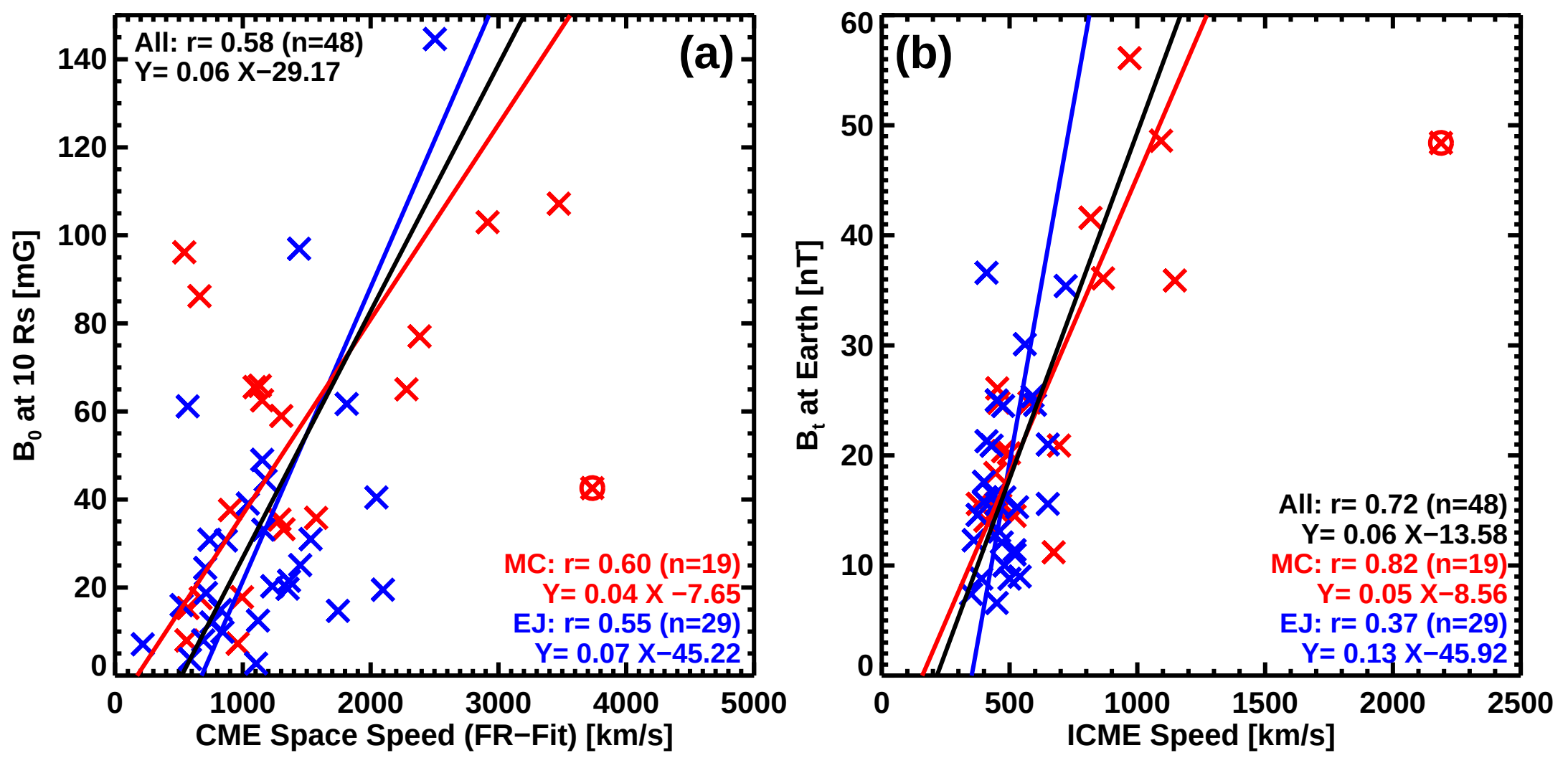

*Figure 8. (a) Scatter plot between the CME speed in the corona obtained from flux-rope (FR) fitting and the field strength of the coronal flux rope at 10 Rs obtained from RC flux and EFR model. The correlations are quite significant for all events (r =0.58), MC subset (r = 0.60), and the EJ subset (0.55). The highest-speed, MC-associated CME (#32, 2001 April 10) is excluded from the correlation (the data point is shown circled) because there is a large discrepancy between the cone-model and FR speeds. The cone model speed is consistent with other CME speeds. Inclusion of this event makes the correlation coefficient of MC events decreases slightly, but remains highly significant. (b) Scatter plot between the observed ICME speed at 1au and the maximum field strength in the ICME interval ($B_t$ listed in Table1). The ICME speeds were obtained from Gopalswamy et al. (2010b). The correlations are statistically significant for all events (r = 0.73), MCs (0.82), and EJs (0.37). The event with the highest MC speed (#45, 2003 October 29) was excluded from the correlation (the data point is shown circled) because of a large discrepancy between observed peak field strength and the Marubashi et al. (2015) fit. Inclusion of this event decreases the correlation coefficient of MCs only slightly to r = 0.74, which is still highly significant.*

The present study provides a physical explanation for the observed correlation between the peak values of speed and magnetic field strength in MCs (Gonzalez et al. 1998; Gopalswamy et al. 2015c). We showed that the peak axial field strength of the coronal flux ropes is proportional to the RC flux, which in turn, is related to the speed of the flux rope. As expected, the space speed and axial field strength of the coronal flux ropes have a strong correlation ($r = 0.58$ with $r_c = 0.240$ for 48 events; see Fig. 8a). In Figure 8b we have shown the scatter plot between ICME speed and its peak field strength similar to the plot published by Gonzalez et al. (1998). We see that the correlation is very high ($r = 0.73$ with $r_c = 0.240$ for 48 events). It must be noted that the range of speeds and magnetic field strengths in Fig. 8b is much larger than that (<700 km/s, <35 nT) used by Gonzalez et al. (1998). Furthermore, the correlation is significant for both MCs and EJs in Fig. 8b, whereas Gonzalez et al. (1998) found no correlation for EJs. There is also good correlation between the CME space speed in Fig.8a and the ICME field strength in Fig. 8b, giving a regression line $B_t = 0.012V_{sp} + 7.0$, with a correlation coefficient of 0.65. We can conclude that a flux rope with higher initial speed (or kinetic energy) is indicative of a flux rope with a larger magnetic content, ultimately stemming from the RC flux. This can also be seen from the correlation between CME space speed and the ICME magnetic field strength

One of the significant findings of this study is that while both MC and EJ type ICMEs have flux rope structure, there are some differences in the flux rope properties. However, the difference is certainly not flux rope (MC) vs. non-rope (EJ) structure as was suggested by some authors (e.g., Gosling 1990). Gosling (1990) suggested that the EJ events may be non-ropes, i.e., the field lines simply expand from the source region without the helical structure. However, such expansion would not involve reconnection and therefore no post eruption arcades or flux ropes would be involved. We have shown that both MC and EJ type ICMEs have similar arcade structure at the Sun and charge states at 1 au implying flare association (Yashiro et al. 2013; Gopalswamy et al. 2013b). We showed that the difference between the MC and EJ source regions is in terms of the amount of RC flux involved, resulting in a weaker magnetic content of the coronal and 1-au flux ropes.

## 7. Conclusions

We have shown that it is possible to fully characterize the geometrical and magnetic properties of CME flux ropes ejected from the Sun soon after the eruption. The flux rope can then be input to global MHD models for prediction purposes. The main conclusions of this study are:

1. In order to obtain the RC flux in an eruption, we need the following information: (i) post-eruption arcades from EUV or soft X-rays and (ii) a photospheric magnetogram taken near the eruption time.
2. A coronal flux rope can be constructed by combining the geometric properties of the flux rope obtained from coronagraph images and the magnetic properties derived from the RC flux under the assumption of a force-free flux rope.
3. The RC flux is well correlated with flare size, flare fluence, CME speed, CME width, CME kinetic energy, and 1-au magnetic field strength of the flux rope associated with the CME.
4. The RC flux in the source regions of MC- and EJ-type ICMEs is significantly different. The flux ropes in EJ events have a smaller magnetic content (lower axial field strength and poloidal flux) compared to those in MCs. This is true near the Sun and at 1 au.
5. The magnetic field strength in coronal flux ropes is higher than that in the ambient corona by a factor of 2.8 for MCs and 1.8 for EJs. This can be tested using direct observations to be obtained by the Solar Probe Plus mission.
6. The magnetic field strength in a coronal flux rope falls faster than the ambient field strength, due to the expansion resulting from higher internal magnetic pressure. This is also reflected in the larger flux rope size at 1 AU in MCs than in EJs.
7. The relation between the RC flux and the poloidal flux of 1-au flux ropes obtained previously for MC events also holds when the EJ events are included.

8. The high correlation between the coronal flux ropes constructed from eruption data and the 1-au flux ropes constructed from in situ data suggests that one can improve the space weather prediction by providing realistic input to global MHD models.

9. The approximate equality between the flare RC flux and the poloidal flux of the flux rope formed due to reconnection, strongly suggests that a pre-eruption flux rope may not exist.

**Author Contributions**

N. Gopalswamy conceived the project, participated in the data analysis and interpretation of the results, and drafted the manuscript. S. Akiyama performed correlation analysis and created some of the figures. S. Yashiro performed data analysis and created some figures. H. Xie fitted flux ropes to the coronagraph data and obtained the flux rope parameters.

**Acknowledgments**: We thank R. L. Moore, S. T. Wu, D. Faulkner, S. Tiwari, M. Jin, N. Lugaz, and Q. Hu for helpful discussions. This work was supported by NASA Heliophysics Guest Investigator program. SOHO is a project of international cooperation between ESA and NASA.

Table 1. List of eruptions with ICME, CME, and Flare parameters

| | ICME | | | | CME | | | | | | Flare | | | | Coronal Flux Rope | | | | | 1-au FR[h] | |
|---|---|---|---|---|---|---|---|---|---|---|---|---|---|---|---|---|---|---|---|---|---|
| Event #[a] | Type[b] | Date yy/mm/dd | Time UT | $B_t$ nT | Date mm/dd | Time UT | Width deg | $V_{sky}$ km/s | $V_{sp}$ km/s[d] | Mass $10^{15}$ g | Location | X-ray Size[f] | Fluence $10^{-2}$ J/m$^2$ | $\Phi_r$ $10^{21}$ Mx | $\Lambda$ | $R_{tip}$ Rs | Direction | $V_{spf}$ km/s[g] | $B_0$ at 10 Rs mG | $\Phi_p$ $10^{21}$ Mx | $B_0$ at 1 au nT |
| 1 | MC | 97/01/10 | 05:18 | -- | 01/06 | 15:10 | 360 | 136 | 232 | 0.58 | S18E06 | A1.1 | ---- | ---- | 0.95 | 5.50 | S18W01 | 258 | ----- | 3.13 | 19.0 |
| 2 | MC | 97/05/15 | 09:06 | 26.1 | 05/12 | 05:30 | 360 | 464 | 749 | 4.16 | N21W08 | C1.3 | 0.25 | 2.4 | 1.00 | 26.00 | N01W02 | 670 | 17.6 | 5.32 | 35.1 |
| 3 | EJ+ | 97/12/11 | 03:45 | 14.6 | 12/06 | 10:27 | 223 | 397 | 592 | 20.1 | N45W10 | B7.0 | 2.16 | 7.7 | 1.10 | 7.80 | N15W30 | 569 | 61.2 | 4.21 | 29.2 |
| 4 | EJ? | 98/05/03 | 19:00 | 10.0 | 05/01 | 23:40 | 360 | 585 | 866 | 6.85 | S18W05 | M1.2 | 1.60 | 3.0 | 0.50 | 22.50 | S01E14 | 826 | 14.9 | ---- | ---- |
| 5 | EJ+ | 98/05/04 | 10:00 | 21.0 | 05/02 | 14:06 | 360 | 938 | 1168 | 7.73 | S15W15 | X1.1 | 6.70 | 3.6 | 0.60 | 28.50 | N08W05 | 2097 | 19.5 | 0.74 | 8.80 |
| 7 | EJ+ | 98/11/07 | 22:00 | 16.2 | 11/04 | 07:54 | 360 | 523 | 809 | 6.99 | N17W01 | C1.6 | ---- | 4.1 | 0.70 | 23.50 | N25W01 | 706 | 24.4 | 5.24 | 42.4 |
| 8 | EJ+ | 98/11/13 | 04:30 | 21.3 | 11/09 | 18:18 | 190 | 325 | 513 | 2.84 | N15W05 | C2.5 | 0.36 | 2.5 | 1.00 | 9.90 | N15W05 | 712 | 18.8 | 5.86 | 26.1 |
| 9 | MC | 99/04/16 | 20:18 | 24.8 | 04/13 | 03:30 | 261 | 291 | 476 | 1.40 | N16E00 | B4.3 | 0.07 | 1.5 | 0.60 | 8.00 | S02W06 | 560 | 8.0 | 5.35 | 28.6 |
| 10 | EJ+ | 99/06/27 | 21:30 | 12.1 | 06/24 | 13:31 | 360 | 975 | 1143 | 2.29 | N29W13 | C4.1 | 3.30 | 5.2 | 0.70 | 7.50 | N25W15 | 1531 | 31.0 | 0.77 | 8.70 |
| 13 | EJ+ | 99/09/22 | 21:00 | 24.6 | 09/20 | 06:06 | 360 | 604 | 900 | 0.89 | S20W05 | C1.4 | ---- | 2.7 | 1.80 | 7.80 | S20W05 | 868 | 30.7 | 3.05 | 25.9 |
| 14 | EJ+ | 99/10/21 | 18:30 | 36.6 | 10/18 | 00:06 | 240 | 144 | 314 | 3.43 | S30E15 | C1.2 | 0.09 | 0.90 | 1.10 | 5.50 | S30E15 | 217 | 7.1 | 2.08 | 38.4 |
| 15 | EJ+ | 00/01/22 | 18:00 | 17.6 | 01/18 | 17:54 | 360 | 739 | 1077 | 6.94 | S19E11 | M3.9 | 5.70 | 7.5 | 0.70 | 6.50 | S10E29 | 1179 | 44.4 | 2.22 | 14.3 |
| 16 | MC | 00/02/21 | 09:48 | 18.4 | 02/17 | 21:30 | 360 | 728 | 973 | 14.7 | S29E07 | M1.3 | 2.70 | 3.0 | 0.70 | 19.50 | S12W02 | 994 | 17.8 | 4.56 | 23.0 |
| 17 | EJ+ | 00/07/11 | 01:30 | 16.4 | 07/07 | 10:26 | 360 | 453 | 766 | 1.82 | N04E00 | C4.5 | 3.45 | 3.9 | 1.10 | 8.50 | S17W05 | 739 | 30.9 | 2.66 | 22.2 |
| 18 | EJ+ | 00/07/11 | 22:48 | 11.4 | 07/08 | 23:50 | 161 | 483 | 687 | 10.0[e] | N18W12 | C4.0 | 2.69 | 7.1 | 0.90 | 8.50 | N18W06 | 1152 | 49.0 | 7.94 | 16.7 |
| 19 | MC | 00/07/15 | 21:06 | 48.6 | 07/14 | 10:54 | 360 | 1674 | 2061 | 13.9 | N22W07 | X5.7 | 75.00 | 13.1 | 0.50 | 6.10 | N18W14 | 2281 | 65.0 | 18.0 | 74.1 |
| 20 | EJ+ | 00/07/27 | 08:28 | 7.4 | 07/23 | 05:30 | 181 | 631 | 850 | 3.08 | S13W05 | ---- | ---- | 1.9 | 0.80 | 7.50 | S13E04 | 1119 | 12.5 | 0.48 | 9.80 |
| 21 | MC | 00/07/28 | 21:06 | 15.6 | 07/25 | 03:30 | 360 | 528 | 900 | 1.63 | N06W08 | M8.0 | 2.80 | 1.1 | 0.80 | 8.50 | S15E04 | 960 | 7.3 | ---- | ---- |
| 23 | MC | 00/08/12 | 06:06 | -- | 08/09 | 16:30 | 360 | 702 | 1007 | 7.02 | N20E12 | ---- | 1.20 | ---- | 1.00 | 6.50 | N17E05 | 1024 | ----- | 9.27 | 34.9 |
| 24 | MC | 00/09/18 | 01:54 | 36.10 | 09/16 | 05:18 | 360 | 1215 | 1568 | 9.59 | N14W07 | M5.9 | 9.80 | 5.5 | 0.80 | 11.80 | N08W07 | 1574 | 35.8 | 6.96 | 37.2 |
| 25 | EJ+ | 00/10/05 | 13:13 | 15.3 | 10/02 | 03:50 | 360 | 525 | 877 | 6.76 | S09E07 | C4.1 | 0.55 | 0.37 | 1.00 | 7.10 | S19E08 | 1104 | 2.7 | 1.38 | 10.5 |
| 26 | MC | 00/10/13 | 18:24 | 14.1 | 10/09 | 23:50 | 360 | 527 | 903 | 14.3 | N01W14 | C6.7 | 1.70 | 5.1 | 0.90 | 7.50 | N20W14 | 1287 | 35.4 | 4.35 | 19.7 |
| 27 | MC | 00/11/06 | 23:06 | 24.8 | 11/03 | 18:26 | 360 | 291 | 701 | 4.85 | N02W02 | C3.2 | 3.22 | 14.9 | 0.80 | 10.10 | N02E05 | 542 | 96.2 | 3.24 | 25.1 |
| 28 | EJ+ | 00/11/27 | 05:00 | 24.5 | 11/24 | 05:30 | 360 | 1289 | 1611 | 6.33 | N20W05 | X2.0 | 8.30 | 2.7 | 0.60 | 6.80 | N30W18 | 1745 | 14.7 | 2.80 | 23.0 |
| 29 | EJ+ | 01/03/04 | 04:00 | 13.1 | 02/28 | 14:50 | 232 | 313 | 500 | 2.41 | S17W05 | B4.2 | 0.12 | 2.7 | 0.70 | 6.50 | S05W15 | 522 | 16.0 | 1.12 | 12.0 |
| 30 | EJ+ | 01/03/22 | 22:30 | 8.8 | 03/19 | 05:26 | 360 | 389 | 653 | 1.19 | S20W00 | PEA | ---- | 1.3 | 0.70 | 7.00 | N05W10 | 691 | 8.0 | 0.14 | 18.5 |
| 31 | EJ+ | 01/04/11 | 22:30 | 35.4 | 04/09 | 15:54 | 360 | 1192 | 1482 | 4.50 | S21W04 | M7.9 | 13.00 | 11.3 | 0.60 | 21.40 | S09W10 | 1813 | 61.7 | ---- | ---- |
| 32 | MC | 01/04/12 | 07:54 | 20.9 | 04/10 | 05:30 | 360 | 2411 | 2940 | 9.12 | S23W09 | X2.3 | 30.00 | 7.2 | 0.70 | 3.40 | S23W05 | 3735 | 42.6 | 1.43 | 44.9 |
| 33 | MC | 01/04/29 | 01:54 | 11.2 | 04/26 | 12:30 | 360 | 1006 | 1257 | 10.0[e] | N20W05 | M1.5 | 5.10 | 12.0 | 0.60 | 8.90 | N20W03 | 1093 | 65.5 | 23.9 | 14.5 |
| 34 | EJ- | 01/08/13 | 07:00 | 12.3 | 08/09 | 10:30 | 175 | 479 | 682 | 3.59 | N11W14 | PEA | ---- | 1.2 | 1.20 | 7.00 | N02W18 | 842 | 9.9 | ---- | ---- |
| 35 | EJ+ | 01/10/12 | 03:30 | 25.4 | 10/09 | 11:30 | 360 | 973 | 1156 | 11.8 | S28E08 | M1.4 | 4.00 | 3.6 | 0.90 | 6.00 | S28E01 | 1449 | 25.1 | 11.9 | 35.8 |
| 36 | MC | 02/03/19 | 22:54 | 15.6 | 03/15 | 23:06 | 360 | 957 | 1297 | 22.5 | S08W03 | M2.2 | 13.00 | 11.5 | 0.60 | 14.10 | N15W01 | 1151 | 62.5 | ---- | ---- |
| 37 | MC | 02/04/18 | 04:18 | 14.5 | 04/15 | 03:50 | 360 | 720 | 1143 | 1.93 | S15W01 | M1.2 | 7.10 | 10.8 | 0.60 | 24.00 | S01W05 | 1302 | 59.0 | 5.63 | 23.7 |
| 38 | EJ+ | 02/05/11 | 13:00 | 20.9 | 05/08 | 13:50 | 360 | 614 | 990 | 0.58 | S12W07 | C4.2 | 1.20 | 2.9 | 0.90 | 6.40 | S12W05 | 1231 | 20.3 | 0.63 | 22.3 |
| 39 | MC | 02/05/19 | 03:54 | 20.4 | 05/16 | 00:50 | 360 | 600 | 823 | 4.25 | S23E15 | C4.5 | 1.50 | 4.7 | 1.10 | 25.10 | S23E05 | 900 | 37.6 | 4.69 | 28.5 |
| 40 | EJ- | 02/05/20 | 11:00 | -- | 05/17 | 01:27 | 45 | 461 | 663 | 1.56 | S20E14 | ---- | ---- | ---- | 1.80 | 7.50 | S28E20 | 743 | ----- | 1.64 | 12.0 |
| 41 | EJ- | 02/05/30 | 07:09 | 8.8 | 05/27 | 13:27 | 161 | 1106 | 1372 | 3.24 | N22E15 | C3.7 | 1.50 | 3.1 | 0.90 | 22.40 | N32E20 | 1362 | 21.5 | ---- | ---- |
| 42 | EJ+ | 02/07/18 | 12:00 | 6.6 | 07/15 | 21:30 | 188 | 1300 | 1586 | 18.7 | N19W01 | M1.8 | 4.30 | 8.2 | 0.50 | 5.80 | N29E15 | 2046 | 40.5 | ---- | ---- |
| 43 | MC | 02/08/01 | 11:54 | 14.9 | 07/29 | 12:07 | 161 | 222[c] | 400 | 8.44 | S10W10 | M4.7 | 8.50 | 10.2 | 0.80 | 7.50 | S02W10 | 1134 | 65.8 | 1.99 | 20.9 |
| 44 | MC | 03/08/18 | 11:36 | 20.2 | 08/14 | 20:06 | 360 | 378 | 766 | 10.0[e] | S10E02 | C3.8 | 1.63 | 10.9 | 1.10 | 8.80 | N12E10 | 662 | 86.2 | 2.67 | 19.8 |
| 45 | MC | 03/10/29 | 08:00 | 48.4 | 10/28 | 11:30 | 360 | 2459 | 3128 | 10.0[e] | S16E08 | X17.2 | 170.00 | 20.8 | 0.50 | 4.10 | S16E20 | 2916 | 103.0 | 23.9 | 69.5 |
| 46 | MC | 03/10/31 | 02:00 | 35.9 | 10/29 | 20:54 | 360 | 2029 | 2628 | 16.6 | S15W02 | X10.0 | 87.00 | 21.6 | 0.50 | 4.00 | S15E05 | 3474 | 107.2 | 7.58 | 40.5 |
| 47 | EJ+ | 04/01/22 | 08:00 | 30.1 | 01/20 | 00:06 | 360 | 965 | 1248 | 10.1 | S13W09 | C5.5 | 4.60 | 14.0 | 0.90 | 7.80 | S25W10 | 1441 | 97.0 | 9.60 | 22.7 |
| 48 | MC | 04/07/24 | 12:48 | -- | 07/22 | 08:30 | 132 | 899 | 1144 | 0.73 | N04E10 | C5.3 | 0.56 | ---- | 0.70 | 7.80 | N06E05 | 1359 | ----- | 5.35 | 38.6 |
| 49 | MC | 04/11/09 | 20:54 | 41.6 | 11/06 | 02:06 | 214 | 1111 | 1378 | 5.96 | N09E05 | M3.6 | 5.50 | 6.7 | 0.50 | 31.80 | N07W00 | 1319 | 33.3 | ---- | ---- |
| 50 | EJ+ | 04/12/12 | 12:00 | 15.8 | 12/08 | 20:26 | 360 | 611 | 1109 | 5.32 | N05W03 | C2.5 | 0.74 | 2.2 | 0.60 | 6.50 | S05W06 | 754 | 12.0 | 4.42 | 20.9 |
| 51 | EJ+ | 05/01/16 | 14:00 | 11.0 | 01/15 | 06:30 | 360 | 2049 | 2701 | 10.0[e] | N16E04 | M8.6 | 29.00 | 20.8 | 0.90 | 6.10 | N25W01 | 2503 | 144.6 | 1.15 | 11.8 |
| 52 | EJ+ | 05/02/18 | 15:00 | 9.0 | 02/13 | 11:06 | 151 | 584 | 798 | 1.10 | S11E09 | C2.7 | 0.22 | 0.56 | 0.85 | 6.10 | S21E19 | 587 | 0.4 | 2.71 | 9.60 |
| 53 | MC | 05/05/15 | 05:42 | 56.1 | 05/13 | 17:12 | 360 | 1689 | 2171 | 10.0[e] | N12E11 | M8.0 | 18.00 | 9.7 | 0.50 | 7.50 | N05E11 | 2384 | 33.3 | 20.3 | 79.0 |
| 54 | MC | 05/05/20 | 07:18 | 15.5 | 05/17 | 03:26 | 273 | 449 | 649 | 10.0[e] | S15W00 | M1.8 | 1.50 | 2.6 | 0.70 | 7.50 | N08W01 | 569 | 15.3 | 2.22 | 19.8 |
| 56 | EJ+ | 05/07/10 | 10:30 | 25.0 | 07/07 | 17:06 | 360 | 683 | 1173 | 10.0[e] | N09E03 | M4.9 | 5.30 | 4.9 | 1.10 | 8.90 | N12E26 | 1040 | 39.0 | ---- | ---- |
| 57 | EJ+ | 05/09/02 | 19:03 | 15.6 | 08/31 | 11:30 | 360 | 825 | 1283 | 2.39 | N13W13 | C2.0 | 1.20 | 4.8 | 0.90 | 24.50 | N08W25 | 1161 | 33.1 | ---- | ---- |
| 58 | EJ+ | 05/09/15 | 14:24 | -- | 09/13 | 20:00 | 360 | 1866 | 2445 | 18.7 | S09E10 | X1.5 | 55.00 | ---- | 0.60 | 5.90 | S29E21 | 2171 | ----- | ---- | ---- |
| 59 | EJ+ | 06/08/20 | 00:00 | 15.6 | 08/16 | 16:30 | 360 | 888 | 1359 | 10.1 | S16W08 | C3.6 | 3.60 | 3.3 | 0.70 | 8.30 | S28W01 | 1351 | 19.7 | ---- | ---- |

[a]List of ICMEs during solar cycle 23 with solar sources near Disk Center (E15° ≤ source longitude ≤ W15°) (from Gopalswamy et al. 2010; 2013).

[b]MC = Magnetic cloud; EJ = Ejecta; the suffix + indicates that it was possible fit a flux rope to the ejecta by adjusting the plasmag boundaries; - indicates it was not possible to fit a flux rope

[c]Incorrect feature might have been tracked to get the speed

[d]Deprojected speed using cone model (for full halo CMEs) or from the empirical relation in Gopalswamy et al. (2015a) for non-halo CMEs

[e] Assumed mass values
[f] EP = Eruptive prominence; PEA = post-eruption arcade
[g] Space speed obtained by tracking the leading edge of flux ropes
[h] 1-au flux rope properties $B_0$ and $R_0$ are from Marubashi et al. (2015); the poloidal flux at 1 au was computed from $B_0$ and $R_0$.

The list contains 54 events out of the original 59 events because
#6, #12, #55 Dropped from the analysis because the revised solar source location fell outside the longitude criterion.
#11 Dropped from the analysis because this is a known “driverless” event.
#22 Dropped from the analysis because of the uncertainty in identifying the solar source; multiple candidate eruptions exist.